\documentclass[aps,prb,reprint,showpacs,superscriptaddress,preprintnumbers]{revtex4-1}
\usepackage{amsmath,txfonts,bm,graphicx,color,hyperref}
\allowdisplaybreaks[4]

\begin{document}
\title{Orbital angular momentum in a nonchiral topological superconductor}
\author{Atsuo Shitade}
\affiliation{Department of Physics, Kyoto University, Kyoto 606-8502, Japan}
\author{Yuki Nagai}
\affiliation{CCSE, Japan Atomic Energy Agency, 178-4-4, Wakashiba, Kashiwa, Chiba, 277-0871, Japan}
\date{\today}
\begin{abstract}
  We investigate the bulk orbital angular momentum in a two-dimensional time-reversal broken topological superconductor
  with the Rashba spin-orbit interaction, the Zeeman interaction, and the $s$-wave pairing potential.
  Prior to the topological phase transition, we find the crossover from $s$ wave to $p$ wave.
  For the large spin-orbit interaction, even in the topological phase, $L_z/N$ does not reach $-1/2$, which is the intrinsic value in chiral $p$-wave superconductors.
  Here $L_z$ and $N$ are the bulk orbital angular momentum and the total number of electrons at zero temperature, respectively.
  Finally, we discuss the effects of nonmagnetic impurities.
\end{abstract}
\maketitle
\section{Introduction} \label{sec:intro}
Topological superconductors (SCs) have been intensively studied
due to their fundamental interest in the physical realization of Majorana fermions~\cite{RevModPhys.83.1057}
and their future application to topological quantum computation~\cite{RevModPhys.80.1083}.
A typical design principle for two-dimensional time-reversal broken topological SCs is to realize a spinless chiral $p$-wave SC~\cite{PhysRevB.61.10267}.
Possible candidates are the surface of a topological insulator~\cite{PhysRevLett.100.096407}
and a quantum anomalous Hall insulator in proximity to $s$-wave SCs~\cite{PhysRevB.82.184516}.
More promising candidates are conventional spin-orbit coupled systems
with the Zeeman interaction and the $s$-wave pairing potential~\cite{PhysRevLett.101.160401,PhysRevLett.103.020401,PhysRevLett.104.040502,PhysRevB.81.125318}.
In fact, their one-dimensional analog was theoretically proposed~\cite{PhysRevLett.105.077001,PhysRevLett.105.177002}
and later experimentally realized in semiconductor nanowires contacted with $s$-wave SCs~\cite{Mourik2012,Rokhinson2012,Das2012,Deng2012}.

In a chiral $p$-wave SC, each Cooper pair carries the orbital angular momentum (AM) $\ell = 1$,
which leads to the Chern number $C = 1$
and the bulk orbital AM $L_z/N = 1/2$ at zero temperature~\cite{Ishikawa01061977,Ishikawa01041980,PhysRevB.21.980,volovik1995,JPSJ.67.216,Goryo1998549,%
PhysRevB.69.184511,PhysRevB.84.214509,PhysRevB.85.100506,PhysRevB.90.134510}.
Here $N$ is the total number of electrons, and we set $\hbar = 1$.
Note that $L_z/N$ is known to deviate from $1/2$ due to retroreflection by edge roughness~\cite{PhysRevB.84.214509}
and the opposite chirality with the vorticity $w_{-1} = +2$ induced around the edge~\cite{PhysRevLett.101.150409,PhysRevA.81.053605}.
However, these effects do not occur without the edge, disorder, or a vortex, and it is natural to call $L_z/N = 1/2$ the intrinsic value in chiral $p$-wave SCs.
In higher-order chiral SCs with $\ell > 1$, the Chern number and the bulk orbital AM are equal to $C = \ell$ and $L_z/N = \ell/2$, respectively~\cite{PhysRevB.90.134510}.
Although it was recently found that the bulk orbital AM is totally suppressed
due to the edge states without the particle-hole symmetry by themselves~\cite{PhysRevLett.114.195301,Volovik2014,PhysRevB.90.224519},
still we can call $L_z/N = \ell/2$ the intrinsic value in chiral SCs.
Thus, the Chern number and the bulk orbital AM appear not to be independent of each other.

Generally speaking, topological invariants are robust against symmetry-preserving perturbations which do not close the gap,
while the orbital AM is not conserved in the presence of spin-orbit interactions (SOIs).
In the specific model with the Rashba SOI $\alpha$, the Zeeman interaction $h$,
and the $s$-wave pairing potential $\Delta$~\cite{PhysRevLett.101.160401,PhysRevLett.103.020401,PhysRevLett.104.040502,PhysRevB.81.125318},
which is dubbed the Rashba+Zeeman+$s$-wave model below,
the Rashba SOI neither closes the gap nor changes the Chern number.
One question is whether the Chern number and the bulk orbital AM are not independent even in the presence of SOIs.
If not, a subsequent question is what the bulk orbital AM characterizes in topological SCs.

One possible answer to the second question is the impurity effects.
In contrast to the Chern number which characterizes a topological SC,
the bulk orbital AM may characterize how close it is to a spinless chiral SC with the same Chern number and how fragile it is to nonmagnetic impurities.
Topological SCs are not always robust against nonmagnetic impurities,
namely, the midgap bound states appear and the critical temperature $T_{\rm c}$ is suppressed~\cite{PhysRevLett.110.020401,JPSJ.83.094722,JPSJ.84.034711}.
These impurity effects in the Rashba+Zeeman+$s$-wave model depend on $h$ and are similar to those in a chiral $p$-wave SC for large $h$.

There is another evidence that the bulk orbital AM is related to the robustness against nonmagnetic impurities.
The impurity effects on the bound state in a single vortex core were investigated
in the Rashba+Zeeman+$s$-wave model~\cite{1742-6596-568-2-022028}.
There are two types of vortices: a parallel (antiparallel) vortex when the signs of the vorticity and the Zeeman interaction are opposite (same).
Note that our sign convention of the Zeeman interaction is opposite to that in Ref.~\onlinecite{1742-6596-568-2-022028}.
The authors found that the low-energy scattering rate is suppressed in the case of an antiparallel vortex compared to the case of a parallel vortex,
and that the robustness of the bound state against nonmagnetic impurities depends on the Rashba SOI.
Thus, the impurity effects may be characterized by the bulk orbital AM rather than the Chern number.

In this paper, we investigate the bulk orbital AM in a two-dimensional time-reversal broken topological SC.
We focus on the Rashba+Zeeman+$s$-wave model~\cite{PhysRevLett.101.160401,PhysRevLett.103.020401,PhysRevLett.104.040502,PhysRevB.81.125318}.
We calculate the bulk orbital AM both by the Berry-phase formula~\cite{PhysRevB.90.134510} and in the circular disk.
One advantage of the Berry-phase formula is that the wave functions in the reciprocal space are simpler compared to those in the real space.
We explain the effects of nonmagnetic impurities
in the absence~\cite{PhysRevLett.110.020401,JPSJ.83.094722,JPSJ.84.034711} and presence~\cite{1742-6596-568-2-022028} of a vortex.

\section{Model and method} \label{sec:model}
\begin{figure*}
  \centering
  \includegraphics[clip,width=0.67\textwidth]{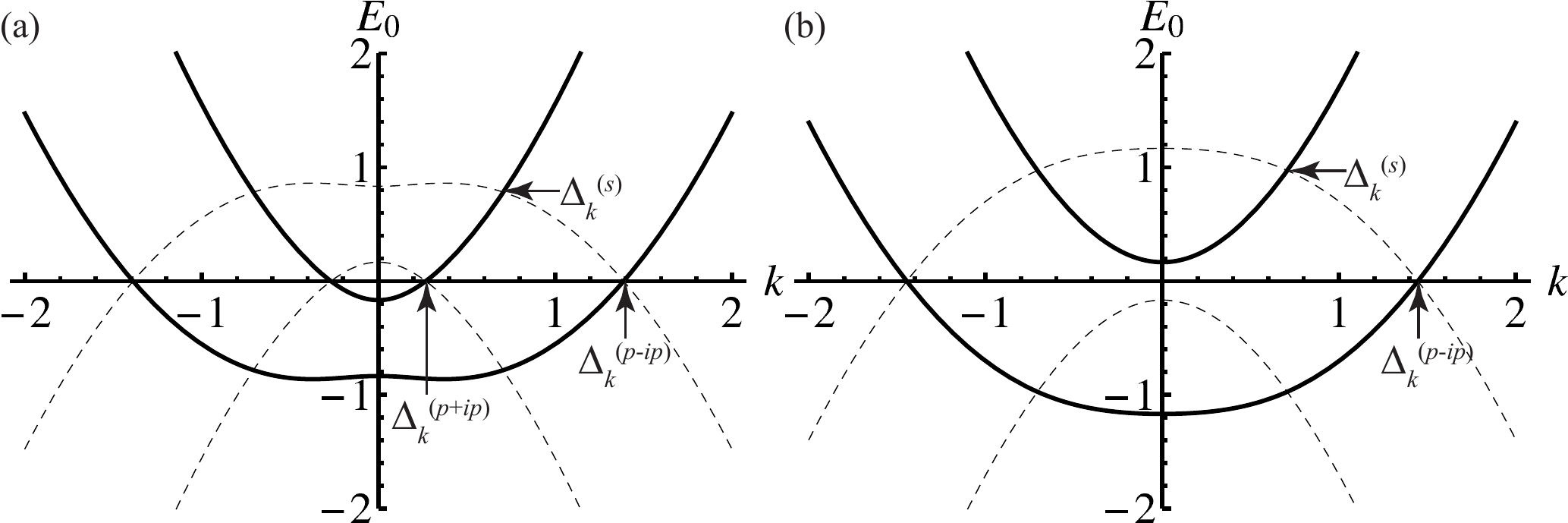}
  \caption{%
  Effective pairing potentials induced by the $s$-wave pairing potential.
  Solid and broken lines indicate the electron and hole bands, respectively.
  (a) For $h = 0.33 < h_{\rm c}$, the pairing potentials $\Delta_{\vec k}^{(p \mp i p)}$ open the gaps in the lower and upper bands, respectively.
  (b) For $h = 0.67 > h_{\rm c}$, the upper band is away from the chemical potential, and $\Delta_{\vec k}^{(p - i p)}$ opens the gap in the lower band.
  We set $\alpha = 1$.
  } \label{fig:e0k}
\end{figure*}

The Rashba+Zeeman+$s$-wave model~\cite{PhysRevLett.101.160401,PhysRevLett.103.020401,PhysRevLett.104.040502,PhysRevB.81.125318} is represented by
\begin{align}
  H
  = & \frac{1}{2} \int \frac{{\rm d}^2 k}{(2 \pi)^2}
  \Psi_{\vec k}^{\dag} [\xi_k \tau_z + \alpha (k_x \sigma_y - k_y \sigma_x) \tau_z \notag \\
  & - h \sigma_z + \Delta \tau_x] \Psi_{\vec k}, \label{eq:razesc}
\end{align}
where $\Psi_{\vec k} = [c_{{\vec k} \uparrow}, c_{{\vec k} \downarrow}, c_{-{\vec k} \downarrow}^{\dag}, -c_{-{\vec k} \uparrow}^{\dag}]^{\rm T}$
is the Nambu spinor, and $\sigma$'s and $\tau$'s are the Pauli matrices for the spin and Nambu spaces, respectively. 
$\xi_k = k^2/2 m - \mu$ is the kinetic term in reference to the chemical potential $\mu$.
There are four dispersions, $\pm E_{k \mp} = \pm \sqrt{\xi_k^2 + (\alpha k)^2 + h^2 + \Delta^2 \mp 2 g_k^2}$
with $g_k^2 = \sqrt{\xi_k^2 (\alpha k)^2 + \xi_k^2 h^2 + h^2 \Delta^2}$.
Therefore, the gap between $\pm E_{k -}$ closes at $k = 0$ for $h_{\rm c} = \sqrt{\mu^2 + \Delta^2}$,
above which this model has the nontrivial Chern number $C = -1$~\cite{PhysRevLett.103.020401,PhysRevLett.104.040502,PhysRevB.81.125318}.
It can be easily understood by representing Eq.~\eqref{eq:razesc} in the band basis~\cite{PhysRevB.81.125318}.
In Fig.~\ref{fig:e0k}(a) for $h < h_{\rm c}$, the gaps open due to the interband $s$-wave pairing potential $\Delta_k^{(s)} = h \Delta/f_k$,
the intraband $(p - i p)$-wave one $\Delta_{\vec k}^{(p - i p)} = i \alpha k e^{-i \phi} \Delta/f_k$ for the lower band $E_{0 k -} = \xi_k - f_k$,
and the intraband $(p + i p)$-wave one $\Delta_{\vec k}^{(p + i p)} = i \alpha k e^{i \phi} \Delta/f_k$ for the upper band $E_{0 k +} = \xi_k + f_k$,
where $f_k = \sqrt{(\alpha k)^2 + h^2}$.
For $h > h_{\rm c}$, the upper band is far away from the chemical potential as shown in Fig.~\ref{fig:e0k}(b),
and the model is mapped onto a $(p - i p)$-wave SC with $C = -1$.
Below we set $m = 0.5$, $\mu =0.5$, and $\Delta = 0.35$, leading to $h_{\rm c} = 0.61$, and change $\alpha$ or $h$.

First, we calculate the bulk orbital AM in the reciprocal space.
The wave functions with the dispersions $+E_{k -}$, $+E_{k +}$, $-E_{k +}$, and $-E_{k -}$ are simply given by
\begin{subequations} \begin{align}
  N_{k -}
  \begin{bmatrix}
    (h^2 - g_k^2 - h E_{k -}) (\xi_k^2 - g_k^2 + \xi_k E_{k -}) \\
    -i \alpha k e^{i \phi} (h - \xi_k) (\xi_k^2 - g_k^2 + \xi_k E_{k -}) \\
    -\Delta (h - \xi_k) (h^2 - g_k^2 - h E_{k -}) \\
    -i \alpha k e^{i \phi} \Delta (h^2 - \xi_k^2)
  \end{bmatrix}
  = &
  \begin{bmatrix}
    u_{k11} \\ i e^{i \phi} u_{k21} \\ -v_{k22} \\ -i e^{i \phi} v_{k12}
  \end{bmatrix}, \label{eq:razescwf1} \\
  N_{k +}
  \begin{bmatrix}
    -i \alpha k e^{-i \phi} (h + \xi_k) (\xi_k^2 + g_k^2 + \xi_k E_{k +}) \\
    (h^2 + g_k^2 + h E_{k +}) (\xi_k^2 + g_k^2 + \xi_k E_{k +}) \\
    i \alpha k e^{-i \phi} \Delta (h^2 - \xi_k^2) \\
    \Delta (h + \xi_k) (h^2 + g_k^2 + h E_{k +})
  \end{bmatrix}
  = &
  \begin{bmatrix}
    -i e^{-i \phi} u_{k12} \\ u_{k22} \\ i e^{-i \phi} v_{k21} \\ -v_{k11} 
  \end{bmatrix}, \label{eq:razescwf2} \\
  N_{k +}
  \begin{bmatrix}
    -\Delta (h + \xi_k) (h^2 + g_k^2 + h E_{k +}) \\
    i \alpha k e^{i \phi} \Delta (h^2 - \xi_k^2) \\
    (h^2 + g_k^2 + h E_{k +}) (\xi_k^2 + g_k^2 + \xi_k E_{k +}) \\
    i \alpha k e^{i \phi} (h + \xi_k) (\xi_k^2 + g_k^2 + \xi_k E_{k +})
  \end{bmatrix}
  = &
  \begin{bmatrix}
   v_{k11} \\ i e^{i \phi} v_{k21} \\ u_{k22} \\ i e^{i \phi} u_{k12}
  \end{bmatrix}, \label{eq:razescwf3} \\
  N_{k -}
  \begin{bmatrix}
    -i \alpha k e^{-i \phi} \Delta (h^2 - \xi_k^2) \\
    \Delta (h - \xi_k) (h^2 - g_k^2 - h E_{k -}) \\
    i \alpha k e^{-i \phi} (h - \xi_k) (\xi_k^2 - g_k^2 + \xi_k E_{k -}) \\
    (h^2 - g_k^2 - h E_{k -}) (\xi_k^2 - g_k^2 + \xi_k E_{k -})
  \end{bmatrix}
  = &
  \begin{bmatrix}
   -i e^{-i \phi} v_{k12} \\ v_{k22} \\  -i e^{-i \phi} u_{k21} \\ u_{k11}
  \end{bmatrix}, \label{eq:razescwf4}
\end{align} \label{eq:razescwf} \end{subequations}
respectively.
Here $N_{k \mp}$ are the normalization constants, and $u_{kij}$ and $v_{kij}$ are real functions of $k$ only.
In the polar coordinate, the Berry curvature for the wave function $|u_{{\vec k} n} \rangle$ is defined by
$\Omega_{{\vec k} n z}
= i (\langle \partial_k u_{{\vec k} n} | \partial_{\phi} u_{{\vec k} n} \rangle - \langle \partial_{\phi} u_{{\vec k} n} | \partial_k u_{{\vec k} n} \rangle)/k$,
and the integrand for the bulk orbital AM is given by
$({\vec A}_{{\vec k} n} \times {\vec k})_z = -i \langle u_{{\vec k} n} | \partial_{\phi} u_{{\vec k} n} \rangle$,
where $A^i_{{\vec k} n} = i \langle u_{{\vec k} n} | \partial_{k_i} u_{{\vec k} n} \rangle$ is the Berry connection~\cite{PhysRevB.90.134510}.
Since the Berry connection can be interpreted as the expectation value of the position operator ${\vec x}$ in the reciprocal space,
the integrand is interpreted as that of ${\vec x} \times {\vec p}$.
The Chern number $C$, the total number of electrons $N$, the total spin $S_z$, and the bulk orbital AM $L_z$ are calculated by
\begin{subequations} \begin{align}
  C
  = & \sum_{ij} (i - j) [(u_{\infty ij}^2 - v_{\infty ij}^2) - (u_{0 ij}^2 - v_{0 ij}^2)], \label{eq:razescc} \\
  N
  = & \int \frac{{\rm d}^2 k}{(2 \pi)^2} \sum_{ij} 2 v_{kij}^2, \label{eq:razescn} \\
  S_z
  = & \int \frac{{\rm d}^2 k}{(2 \pi)^2} \sum_j (v_{k1j}^2 - v_{k2j}^2), \label{eq:razescsz} \\
  L_z
  = & -\int \frac{{\rm d}^2 k}{(2 \pi)^2} \sum_{ij} (i - j) (u_{kij}^2 - v_{kij}^2), \label{eq:razesclz}
\end{align} \label{eq:razescphys} \end{subequations}
respectively.
Note that the total number of electrons $N$ as well as the bulk orbital AM $L_z$ changes by changing $\alpha$ or $h$ with $\mu$ being fixed.
Nonetheless, the bulk orbital AM per electron $L_z/N$ plays an important role when we discuss the impurity effects below.

The bulk orbital AM can be calculated not only in the reciprocal space but in the real space. 
We consider the circular disk with the radius $r_{\rm c}$. 
The Bogoliubov-de Gennes (BdG) equations in the real space are given as 
\begin{equation}
  [H_0(- i {\bm \nabla}) \tau_z + \Delta \tau_x]
  \begin{bmatrix}
   {\bm u}({\bm r}) \\
   {\bm v}({\bm r})
  \end{bmatrix}
  = E
  \begin{bmatrix}
    {\bm u}({\bm r}) \\
    {\bm v}({\bm r})
  \end{bmatrix},
\end{equation}
where
\begin{equation}
  H_0(- i {\bm \nabla})
  =
  \begin{bmatrix}
    \xi(r, \theta) - h & -\alpha L_-(r, \theta) \\
    \alpha L_+(r, \theta) & \xi(r, \theta) + h 
  \end{bmatrix},
\end{equation}
with $\xi(r, \theta) = -(\partial^2/\partial r^2 + (1/r) \partial/\partial r + (1/r^2) \partial^2/\partial \theta^2)/2 m - \mu$
and $L_{\pm} = e^{\pm i \theta} (\partial/\partial r \pm (i/r) \partial/\partial \theta)$.
With the use of the rotational symmetry along the $z$ axis in the circular disk, the solutions can be expressed as 
\begin{equation}
  \begin{bmatrix}
    {\bm u}_{n}({\bm r}) \\
    {\bm v}_{n}({\bm r})
  \end{bmatrix}
  = e^{i n \theta}
  \begin{bmatrix}
    u_{\uparrow}(r) \\
    e^{i \theta} u_{\downarrow}(r) \\
    v_{\downarrow}(r) \\
    -e^{i \theta} v_{\uparrow}(r) 
  \end{bmatrix}.
\end{equation}
Here $n$ is the quantum number. 
When $[{\bm u}_n({\bm r})^{\rm T}, {\bm v}_n({\bm r})^{\rm T}]$ is a solution with the energy $E_n$,
$[{\bm u}^-_{-n-1}({\bm r})^{\rm T}, {\bm v}^-_{-n-1}({\bm r})^{\rm T}] \equiv [- i \sigma_y {\bm v}_n({\bm r})^{\rm T}, i \sigma_y {\bm u}_n({\bm r})^{\rm T}]$
is the solution with the energy $-E_n$.
We use the normalization condition,
\begin{equation}
  \sum_{\sigma} \int_0^{r_{\rm c}} \int_0^{2 \pi} r {\rm d} r {\rm d} \theta (|u_{\sigma}(r)|^2 + |v_{\sigma}(r)|^2)
  = 1,
\end{equation}
and the boundary conditions,
\begin{subequations} \begin{align}
  \frac{\partial}{\partial r}
  \left.\begin{bmatrix}
    u_{\uparrow}(r) \\
    u_{\downarrow}(r) \\
    v_{\downarrow}(r) \\
    -v_{\uparrow}(r) 
  \end{bmatrix} \right|_{r = 0}
  = & 0, \\
  \begin{bmatrix}
    u_{\uparrow}(r_{\rm c}) \\
    u_{\downarrow}(r_{\rm c}) \\
    v_{\downarrow}(r_{\rm c}) \\
    -v_{\uparrow}(r_{\rm c}) 
  \end{bmatrix}
  = & 0.
\end{align} \end{subequations}
The bulk orbital AM in the real space is expressed as 
\begin{align}
  L_z
  = & 2 \pi \sum_l \sum_{n=0}^{n_{\rm c}} \int_0^{r_{\rm c}} r {\rm d} r [(n |u_{\uparrow}^l(r)|^2 \notag \\
  & + (n + 1) |u_{\downarrow}^l(r)|^2) f(E_l^n) - ((n+1) |v_{\uparrow}^l(r)|^2 \notag \\
  & + n |v_{\downarrow}^l(r)|^2) f(-E_l^n)],
\end{align}
with the $l$-th eigenvalue $E_l^n$ with fixed $n$ and the Fermi-Dirac distribution function $f(\omega) \equiv 1/(e^{\omega/T} + 1)$.
Note that we focus on $T = 0$. 
In order to solve the radial BdG equations, the second-order finite difference method with $N_{\rm g}$ real-space grid points is used. 
Thus, the BdG equations become the $2 N_{\rm g} \times 2 N_{\rm g}$ matrix eigenvalue equations. 
We set the number of the grid points $N_{\rm g} = 1000$, the cutoff $n_{\rm c} = 512 - 1$, and the disk radius $r_{\rm c} = 100$. 
The unit of the real space is defined by the lattice spacing of the tight-binding model in Ref.~\onlinecite{JPSJ.84.034711}.

\section{Results} \label{sec:result}
\begin{figure*}
  \centering
  \includegraphics[clip,width=0.67\textwidth]{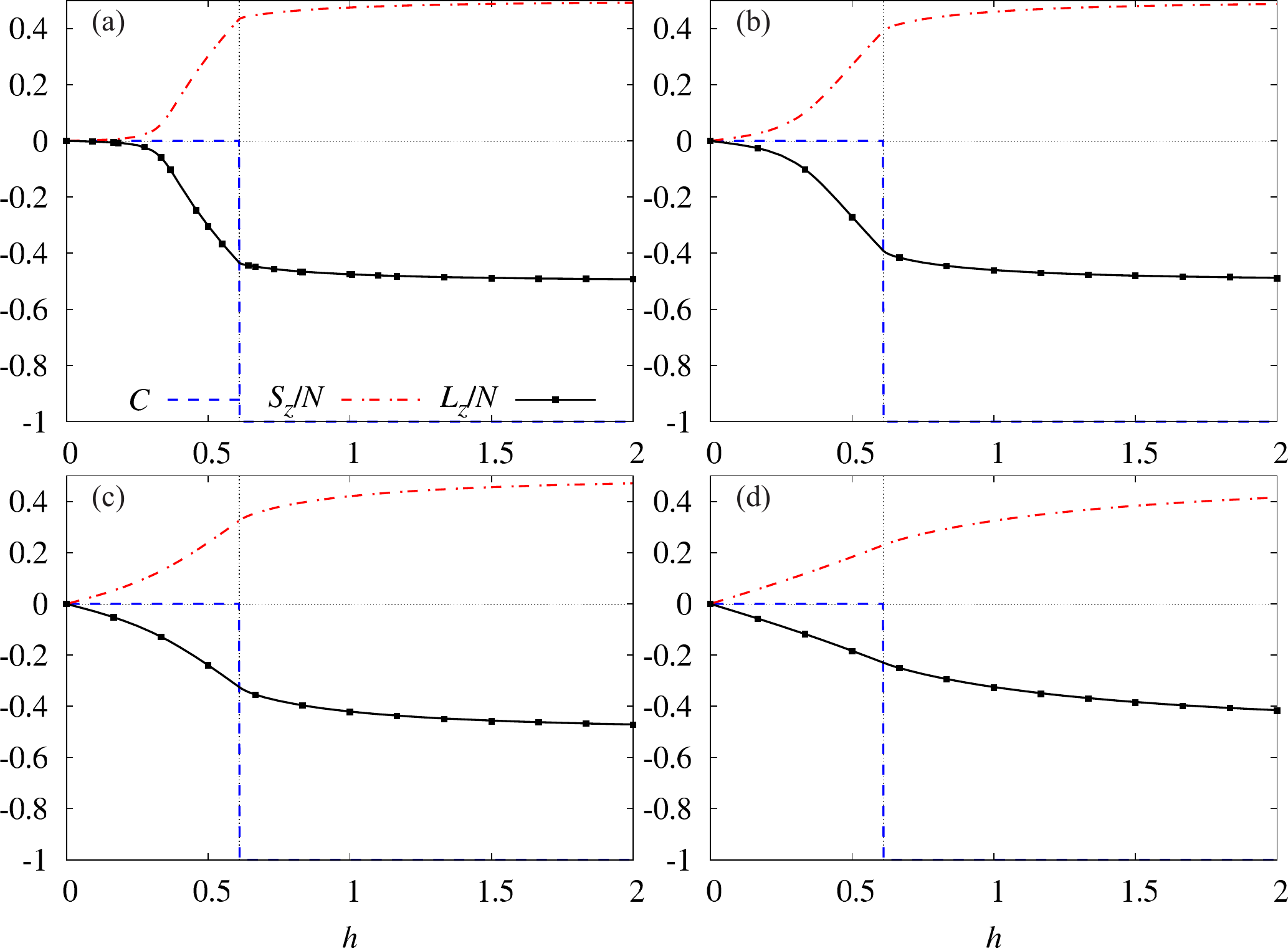}
  \caption{%
  (Color online) $h$ dependence of the Chern number $C$ (blue broken line), the total spin per electron $S_z/N$ (red dashed line),
  and the bulk orbital AM per electron $L_z/N$ (black solid line).
  Black filled squares indicate $L_z/N$ obtained by numerical calculations in the circular disk.
  We set (a) $\alpha = 0.1$, (b) $\alpha = 0.25$, (c) $\alpha = 0.5$, and (d) $\alpha = 1$.%
  } \label{fig:am1}
\end{figure*}

\begin{figure*}
  \centering
  \includegraphics[clip,width=0.67\textwidth]{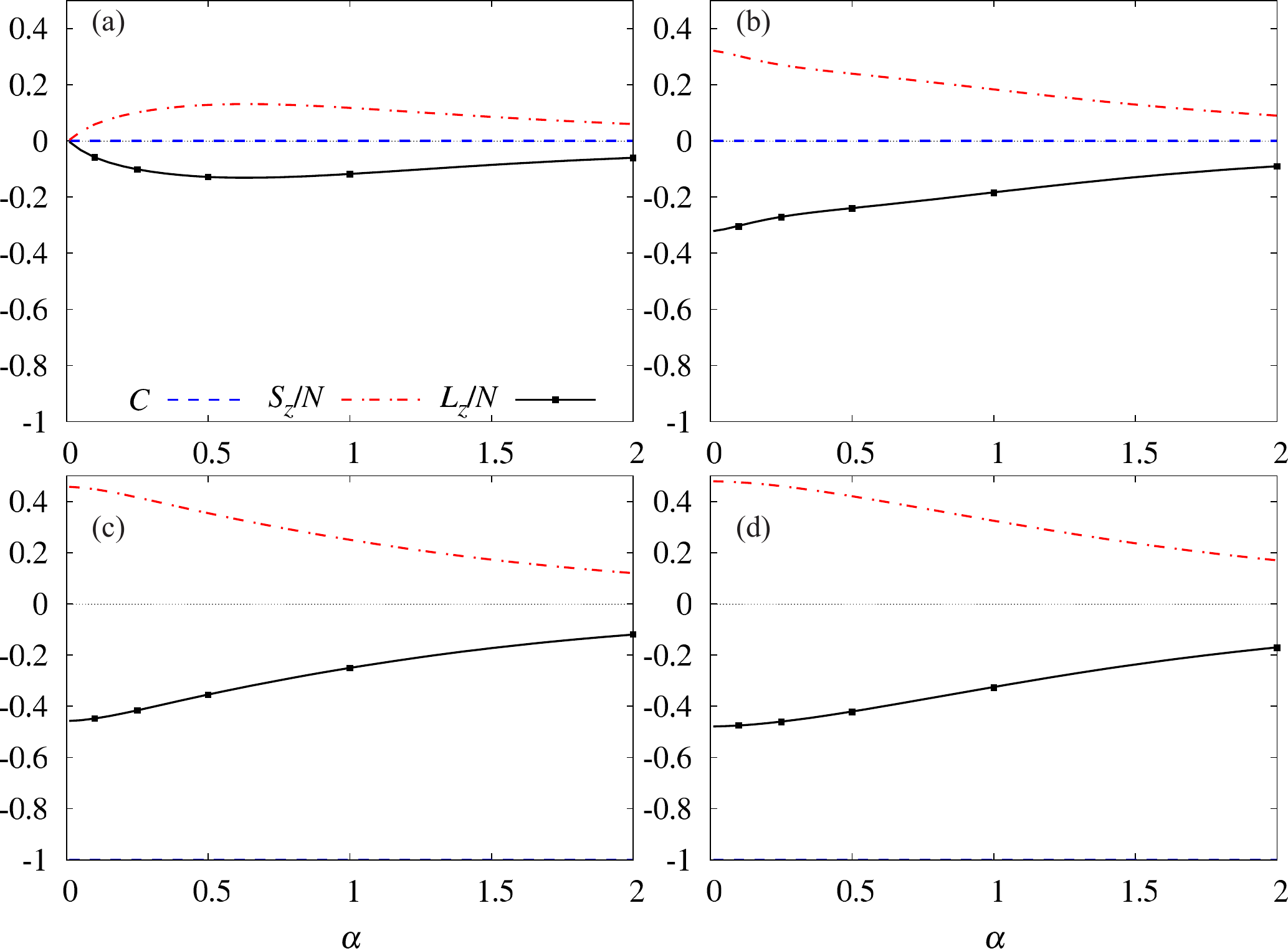}
  \caption{%
  (Color online) $\alpha$ dependence of $C$, $S_z/N$, and $L_z/N$.
  The legends are the same as in Fig.~\ref{fig:am1}.
  We set (a) $h = 0.33 < \Delta$, (b) $\Delta < h = 0.5 < h_{\rm c}$, (c) $h = 0.67 > h_{\rm c}$, and (d) $h = 1 > h_{\rm c}$.%
  } \label{fig:am2}
\end{figure*}

Figures~\ref{fig:am1} and \ref{fig:am2} show the $h$ and $\alpha$ dependences of $C$, $S_z/N$, and $L_z/N$.
$L_z/N$ obtained by the Berry-phase formula completely coincides with that in the circular disk.
For small $\alpha$, we obtain $L_z/N \simeq 0$ for $h < \Delta$ and $L_z/N \simeq -1/2$ for $h > h_{\rm c}$ as shown in Fig.~\ref{fig:am1}(a).
These are expected from the $s$- and $(p - i p)$-wave behaviors, respectively.
Remarkably, in the intermediate region $\Delta < h < h_{\rm c}$, we obtain the nonzero $L_z/N$.
In contrast to $C$, $L_z/N$ does not jump but shows the crossover.
As $\alpha$ increases, the region with $L_z/N \simeq 0$ gets narrower.
For large $\alpha$, as shown in Figs.~\ref{fig:am1}(c) and \ref{fig:am1}(d), $L_z/N$ never goes to $-1/2$ even in the topological phase.
In Fig.~\ref{fig:am2}, we find that $\alpha$ tends to suppress $L_z/N$.
Although the Rashba+Zeeman+$s$-wave model is mapped onto a chiral $p$-wave SC, these two are different in terms of the bulk orbital AM.
These results are summarized in Tab.~\ref{tab:am}.
\begin{table}
  \centering
  \caption{%
  Summary of the $h$ and $\alpha$ dependences of the bulk orbital AM per electron $L_{\rm z}/N$.
  For small $\alpha$, $h^{\ast} = \Delta$.%
  }
  \begin{tabular}{cccc} \hline \hline
    & Small $\alpha$ & Large $\alpha$ & Chern number $C$ \\ \hline
    $h < h^{\ast}$ & $0$ & Finite & $0$ \\
    $h^{\ast} < h < h_{\rm c}$ & Finite & Finite & $0$ \\
    $h_{\rm c} < h$ & $-1/2$ & Finite ($\not= -1/2$) & $-1$ \\ \hline \hline
  \end{tabular} \label{tab:am}
\end{table}

\begin{figure*}
  \centering
  \includegraphics[clip,width=0.67\textwidth]{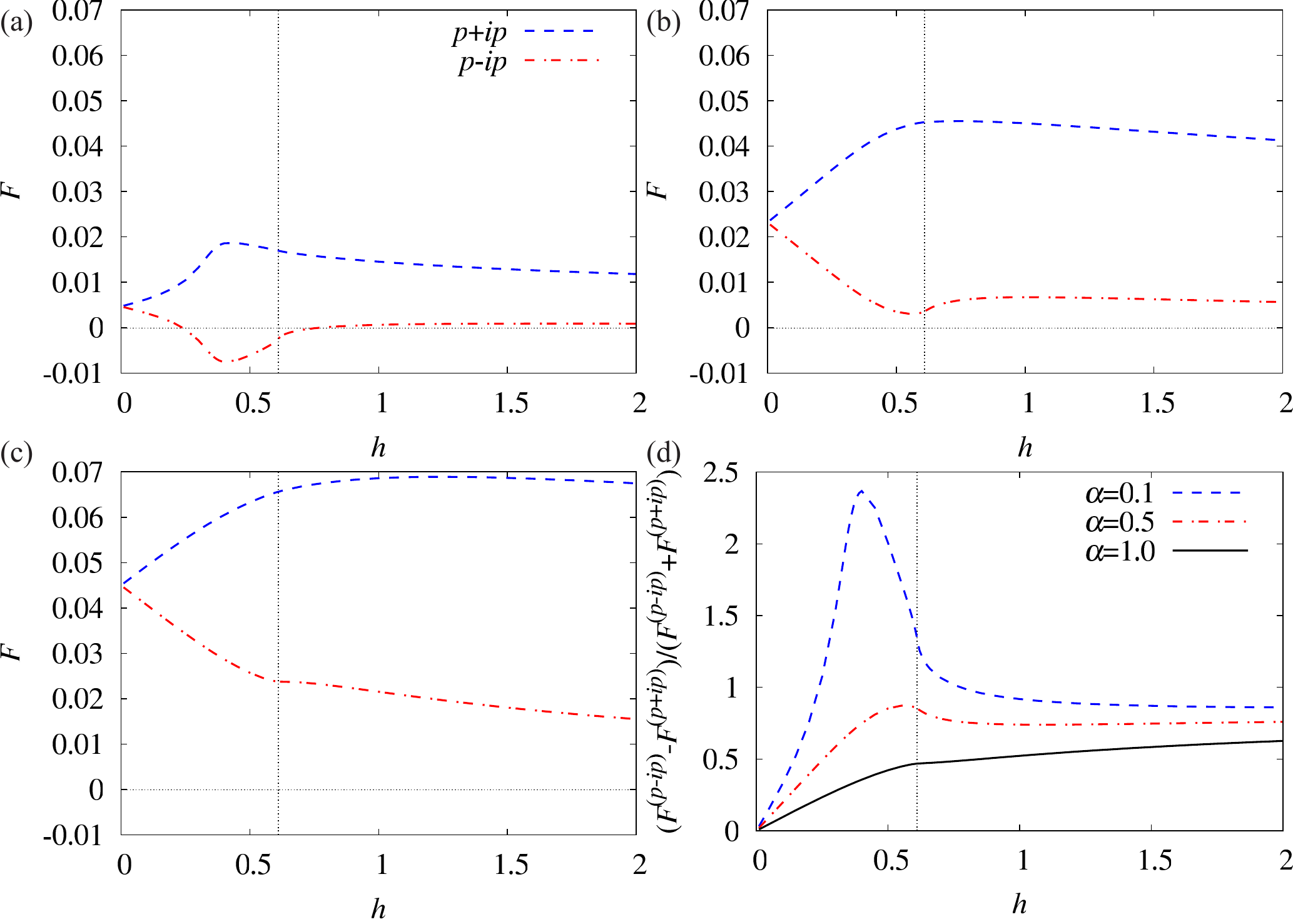}
  \caption{%
  (Color online) (a-c) $h$ dependence of the pairing amplitudes $F^{(p - i p)}$ (blue broken line) and $F^{(p + i p)}$ (red dashed line).
  We set (a) $\alpha = 0.1$, (b) $\alpha = 0.5$, and (c) $\alpha = 1$.
  (d) $h$ dependence of the normalized difference $(F^{(p - i p)} - F^{(p + i p)})/(F^{(p - i p)} + F^{(p + i p)})$ for different $\alpha$.%
  } \label{fig:amp}
\end{figure*}

Let us discuss the origins of the $h$ and $\alpha$ dependences of $L_z/N$.
In the band basis, the interband $s$-wave pairing potential $\Delta_k^{(s)}$ opens the gap
at $k = \sqrt{2 m \mu}$ and $E_{0 k +} = -E_{0 k -} = \sqrt{2 m \alpha^2 \mu + h^2}$ as shown in Fig.~\ref{fig:e0k}.
Therefore, $s$ wave is dominant for $E_{0 k +} < \Delta$, namely, $h < h^{\ast} = \sqrt{\Delta^2 - 2 m \alpha^2 \mu}$.
In the case of small $\alpha$, we obtain $L_z/N \simeq 0$ for $h < \Delta$.
To see this, we evaluate the triplet pairing amplitudes
\begin{subequations} \begin{align}
  \langle c_{{\vec k} \uparrow} c_{{\vec k} \uparrow} \rangle
  = & -i e^{-i \phi} (u_{k11} v_{k12} + u_{k12} v_{k11}), \label{eq:ampup} \\
  \langle c_{{\vec k} \downarrow} c_{{\vec k} \downarrow} \rangle
  = & -i e^{i \phi} (u_{k21} v_{k22} + u_{k22} v_{k21}), \label{eq:ampdn}
\end{align} \label{eq:amp} \end{subequations}
whose pairing symmetries are $(p \mp i p)$ wave.
Figures~\ref{fig:amp}(a-c) show the $h$ dependence of the integrated pairing amplitudes
\begin{subequations} \begin{align}
  F^{(p - i p)}
  = & \int \frac{{\rm d}^2 k}{(2 \pi)^2} (u_{k11} v_{k12} + u_{k12} v_{k11}), \label{eq:iampup} \\
  F^{(p + i p)}
  = & \int \frac{{\rm d}^2 k}{(2 \pi)^2} (u_{k21} v_{k22} + u_{k22} v_{k21}), \label{eq:iampdn}
\end{align} \label{eq:iamp} \end{subequations}
for different $\alpha$.
For $h < h^{\ast}$, both $F^{(p \mp i p)}$ are suppressed as seen in Fig.~\ref{fig:amp}(a).

On the other hand, for $h^{\ast} < h < h_{\rm c}$, $s$ wave is no longer dominant  although it is topologically trivial.
In Figs.~\ref{fig:amp}(a-c), the $(p - i p)$-wave pairing amplitudes $F^{(p - i p)}$ increase.
Owing to the intraband $(p \mp i p)$-wave pairing potentials, the lower and upper bands acquire the orbital AM $\pm \langle \ell_z \rangle$ per electron, respectively.
When the total numbers of electrons in the lower and upper bands are denoted by $N_{\mp}$, respectively,
the bulk orbital AM is represented by $L_z = \langle \ell_z \rangle (N_- - N_+)$,
and hence $L_z/N = \langle \ell_z \rangle (N_- - N_+)/(N_- + N_+)$ negatively increases as a function of $h$.
Note that $N_- > N_+$ and $\langle \ell_z \rangle < 0$.
Especially, in the case of small $\alpha$ and $\Delta$,
we obtain $L_z/N = \langle \ell_z \rangle h/\mu$ because of $N^{\mp} = m (\mu \pm h) \theta(\mu \pm h)/2 \pi$.

For $h > h_{\rm c}$, since the upper band goes away from the chemical potential,
we obtain $N_+ = 0$ and $L_z/N = \langle \ell_z \rangle$.
However, it is not correct to expect $\langle \ell_z \rangle = -1/2$ just because the pairing potential is $(p - i p)$ wave.
In the presence of the Rashba SOI,
not the orbital AM $\ell_z = -i \partial_{\phi}$ but the total AM $j_z = \ell_z + \sigma_z/2$ is conserved, i.e., $[j_z, H_{\vec k}] = 0$.
We find that the wave functions Eqs.~\eqref{eq:razescwf3} and \eqref{eq:razescwf4} are the spontaneous eigenstates of $j_z$ whose eigenvalues are $\pm 1/2$.
Therefore, the bulk total AM $J_z = L_z + S_z$ always vanishes as seen in Figs.~\ref{fig:am1} and \ref{fig:am2}.
For small $\alpha$, spins are polarized  in the $z$ direction owing to the Zeeman interaction, which leads to $\langle \ell_z \rangle \simeq -1/2$.
Correspondingly, the $(p + i p)$-wave pairing amplitude $F^{(p + i p)}$ is almost suppressed in Figs.~\ref{fig:amp}(a,b).
As $\alpha$ increases, spins are tilted, and $\langle \ell_z \rangle$ goes to zero.
Although $F^{(p + i p)}$ itself is no longer suppressed in Fig.~\ref{fig:amp}(c),
we find that the normalized difference between the pairing amplitudes $(F^{(p - i p)} - F^{(p + i p)})/(F^{(p - i p)} + F^{(p + i p)})$ decreases with $\alpha$ increasing
as seen in Fig.~\ref{fig:amp}(d).

\section{Discussion} \label{sec:discussion}
Here we discuss the effects of nonmagnetic impurities in time-reversal-broken topological SCs.
In the topological phase, a strong nonmagnetic impurity behaves like a vortex core, which gives rise to the midgap bound state~\cite{PhysRevLett.110.020401}.
Since a vortex carries the orbital AM, such behavior seems plausible in the topological phase with the bulk orbital AM.
The ratio $|E_{\rm b}/E_{\rm g}|$ monotonically decreases as $h$ increases~\cite{JPSJ.84.034711},
where $E_{\rm b}$ and $E_{\rm g}$ represent the energy level of the bound state and the energy gap, respectively.
This nonuniversal impurity effect can be characterized by the nonuniversal bulk orbital AM.

\begin{figure*}
  \centering
  \includegraphics[clip,width=0.67\textwidth]{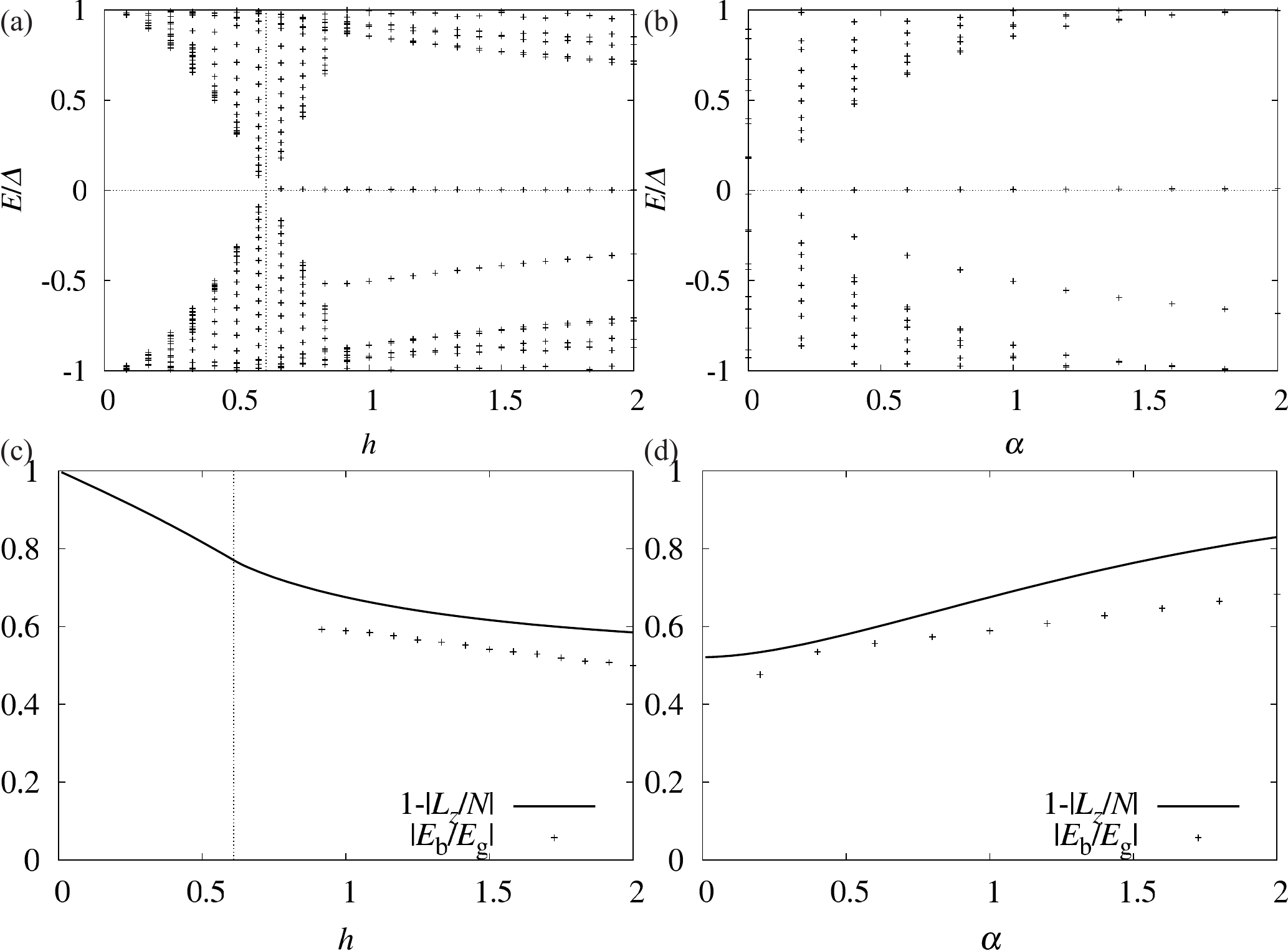}
  \caption{%
  (a) $h$ and (b) $\alpha$ dependences of the eigenvalues $E_l^{n = 0}$ for the BdG equations with a single impurity.
  (c) $h$ and (d) $\alpha$ dependences of the bulk orbital AM per electron $1 - |L_z/N|$ (solid line)
  and the energy level of the midgap bound state $|E_{\rm b}/E_{\rm g}|$ (cross).
  We set (a,c) $\alpha = 1$ and (b,d) $h = 1$.%
  } \label{fig:bound}
\end{figure*}

To check this conjecture, we introduce a single impurity potential $V_0 \theta(r_{\rm i} - r)$ at the center of the circular disk and solve the BdG equations.
Here $\theta(x)$ is the step function,
and we set the radius of the impurity potential $r_{\rm i} = 0.5$ and the strength $V_0 = 100$.
Remember that we set the number of the grid points $N_{\rm g} = 1000$ and the disk radius $r_{\rm c} = 100$.
Figures~\ref{fig:bound}(a,b) show the $h$ dependence of the eigenvalues $E_l^{n = 0}$ for $\alpha = 1$ and $\alpha$ dependence for $h = 1$, respectively.
We can clearly see the midgap bound state in addition to the edge state $E = 0$ for $h > h_{\rm c}$.
Since the impurity strength is sufficiently large, this bound state is what is called the ``universal midgap bound state'' in Ref.~\onlinecite{PhysRevLett.110.020401}.
Apparently, the obtained midgap bound state is not universal.
The ``universal midgap bound state'' $E \simeq \Delta_0^2/E_{\rm F}$ in a topological SC~\cite{PhysRevLett.110.020401} relied on
the edge state $E(n)  = -(n+1/2) \Delta_0/k_{\rm F} \xi$ in a chiral $p$-wave SC~\cite{PhysRevB.69.184511}.
The former $\Delta_0$ is the $s$-wave pairing potential, but the latter is the chiral $p$-wave one.
We should replace $\Delta_0$ by $\Delta_{\rm eff}$, which depends on $\alpha$ and $h$.
However, we cannot simply interpret $\Delta_{\rm eff}$ as the $(p - i p)$-wave pairing potential
because there coexists the $(p + i p)$-wave component as shown in Fig.~\ref{fig:amp}.
Although we assume the uniform pairing potential and do not solve the BdG equations self-consistently, 
we find that the energy level of this bound state $|E_{\rm b}/E_{\rm g}|$ and the bulk orbital AM $1 - |L_z/N|$ are well correlated
as seen in Figs~\ref{fig:bound}(c,d).
A self-consistent calculation may improve this correlation but is a future issue.

With the appearance of the midgap bound states, the reduction of the critical temperature $T_{\rm c}$ also becomes relevant~\cite{JPSJ.83.094722}.
This crossover of the robustness against nonmagnetic impurities is consistent with the crossover of the bulk orbital AM we have found.
Moreover, we expect that the Rashba SOI makes the topological SC robust because it suppresses the bulk orbital AM.

We can also explain the vorticity-dependent impurity effects on the vortex bound state~\cite{1742-6596-568-2-022028}.
As already mentioned in Sec.~\ref{sec:intro},
the bound state in an antiparallel vortex is more robust against nonmagnetic impurities than that in a parallel vortex.
Furthermore, the low-energy scattering rate in the case of an antiparallel vortex is suppressed for small $\alpha$ as in chiral $p$-wave SCs,
while it is similar to that in the case of a parallel vortex for large $\alpha$ as in $s$-wave SCs.
The orbital AM carried by the antiparallel vortex is almost canceled by the intrinsic orbital AM for small $\alpha$, but not for large $\alpha$.
For sufficiently large $\alpha$, the intrinsic orbital AM is totally suppressed, and two types of vortices cannot be distinguished.
Therefore, we conclude that the impurity effects are characterized not by the integer Chern number but by the nonuniversal bulk orbital AM.

Although it may be difficult to measure the bulk orbital AM directly, there are two alternatives.
One is the Hall viscosity,
which is equal to half the bulk orbital AM in two-dimensional gapped systems at zero temperature~\cite{PhysRevB.79.045308,PhysRevB.84.085316}.
The other is the spin magnetization, whose absolute value is equal to that of the bulk orbital AM but sign is opposite as discussed above.
Especially, in Fig.~\ref{fig:am1}, the slope of $S_z/N$ at $h \to 0$ is nothing but the spin susceptibility $\chi_{zz} = \partial S_z/N \partial h$.
In the presence of the SOI, $\chi_{zz}$ is nonzero due to the interband Van Vleck spin susceptibility.
This is an indirect but simple way to measure the bulk orbital AM.

\section{Summary} \label{sec:summary}
To summarize, we have calculated the bulk orbital AM in the two-dimensional Rashba+Zeeman+$s$-wave topological SC.
The Berry-phase formula coincides with numerical calculations in the circular disk.
Prior to the topological phase transition at $h = h_{\rm c}$,
we have found the crossover from $s$ wave to $p$ wave at $h = h^{\ast}$, where $L_z/N$ starts to increase negatively.
We have also found that the Rashba SOI tends to suppress $S_z/N$ and $L_z/N$ because the bulk total AM always vanishes.
For large $\alpha$, even in the topological phase, $L_z/N$ does not reach $-1/2$, which is the intrinsic value in chiral $p$-wave SCs.
We have explained the impurity effects on the topological SC: the appearance of the midgap bound states~\cite{PhysRevLett.110.020401,JPSJ.84.034711},
the reduction of the critical temperature~\cite{JPSJ.83.094722}, and the vorticity dependence of the vortex bound state~\cite{1742-6596-568-2-022028}.

\begin{acknowledgments}
We thank Y. Ota for helpful discussions and comments. 
The calculations have been performed using the supercomputing system PRIMERGY BX900 at the Japan Atomic Energy Agency. 
A.~S. was supported by a Grant-in-Aid for Japan Society for the Promotion of Science (JSPS) Fellows, Grant No.~$24$-$600$,
and Y.~N. was partially supported by JSPS KAKENHI Grant No.~$26800197$. 
\end{acknowledgments}
%

\begin{thebibliography}{37}%
\makeatletter
\providecommand \@ifxundefined [1]{%
 \@ifx{#1\undefined}
}%
\providecommand \@ifnum [1]{%
 \ifnum #1\expandafter \@firstoftwo
 \else \expandafter \@secondoftwo
 \fi
}%
\providecommand \@ifx [1]{%
 \ifx #1\expandafter \@firstoftwo
 \else \expandafter \@secondoftwo
 \fi
}%
\providecommand \natexlab [1]{#1}%
\providecommand \enquote  [1]{``#1''}%
\providecommand \bibnamefont  [1]{#1}%
\providecommand \bibfnamefont [1]{#1}%
\providecommand \citenamefont [1]{#1}%
\providecommand \href@noop [0]{\@secondoftwo}%
\providecommand \href [0]{\begingroup \@sanitize@url \@href}%
\providecommand \@href[1]{\@@startlink{#1}\@@href}%
\providecommand \@@href[1]{\endgroup#1\@@endlink}%
\providecommand \@sanitize@url [0]{\catcode `\\12\catcode `\$12\catcode
  `\&12\catcode `\#12\catcode `\^12\catcode `\_12\catcode `\%12\relax}%
\providecommand \@@startlink[1]{}%
\providecommand \@@endlink[0]{}%
\providecommand \url  [0]{\begingroup\@sanitize@url \@url }%
\providecommand \@url [1]{\endgroup\@href {#1}{\urlprefix }}%
\providecommand \urlprefix  [0]{URL }%
\providecommand \Eprint [0]{\href }%
\providecommand \doibase [0]{http://dx.doi.org/}%
\providecommand \selectlanguage [0]{\@gobble}%
\providecommand \bibinfo  [0]{\@secondoftwo}%
\providecommand \bibfield  [0]{\@secondoftwo}%
\providecommand \translation [1]{[#1]}%
\providecommand \BibitemOpen [0]{}%
\providecommand \bibitemStop [0]{}%
\providecommand \bibitemNoStop [0]{.\EOS\space}%
\providecommand \EOS [0]{\spacefactor3000\relax}%
\providecommand \BibitemShut  [1]{\csname bibitem#1\endcsname}%
\let\auto@bib@innerbib\@empty
\bibitem [{\citenamefont {Qi}\ and\ \citenamefont
  {Zhang}(2011)}]{RevModPhys.83.1057}%
  \BibitemOpen
  \bibfield  {author} {\bibinfo {author} {\bibfnamefont {X.-L.}\ \bibnamefont
  {Qi}}\ and\ \bibinfo {author} {\bibfnamefont {S.-C.}\ \bibnamefont {Zhang}},\
  }\href {\doibase 10.1103/RevModPhys.83.1057} {\bibfield  {journal} {\bibinfo
  {journal} {Rev. Mod. Phys.}\ }\textbf {\bibinfo {volume} {83}},\ \bibinfo
  {pages} {1057} (\bibinfo {year} {2011})} \BibitemShut {NoStop}%
\bibitem [{\citenamefont {Nayak}\ \emph {et~al.}(2008)\citenamefont {Nayak},
  \citenamefont {Simon}, \citenamefont {Stern}, \citenamefont {Freedman},\ and\
  \citenamefont {Das~Sarma}}]{RevModPhys.80.1083}%
  \BibitemOpen
  \bibfield  {author} {\bibinfo {author} {\bibfnamefont {C.}~\bibnamefont
  {Nayak}}, \bibinfo {author} {\bibfnamefont {S.~H.}\ \bibnamefont {Simon}},
  \bibinfo {author} {\bibfnamefont {A.}~\bibnamefont {Stern}}, \bibinfo
  {author} {\bibfnamefont {M.}~\bibnamefont {Freedman}}, \ and\ \bibinfo
  {author} {\bibfnamefont {S.}~\bibnamefont {Das~Sarma}},\ }\href {\doibase
  10.1103/RevModPhys.80.1083} {\bibfield  {journal} {\bibinfo  {journal} {Rev.
  Mod. Phys.}\ }\textbf {\bibinfo {volume} {80}},\ \bibinfo {pages} {1083}
  (\bibinfo {year} {2008})} \BibitemShut {NoStop}%
\bibitem [{\citenamefont {Read}\ and\ \citenamefont
  {Green}(2000)}]{PhysRevB.61.10267}%
  \BibitemOpen
  \bibfield  {author} {\bibinfo {author} {\bibfnamefont {N.}~\bibnamefont
  {Read}}\ and\ \bibinfo {author} {\bibfnamefont {D.}~\bibnamefont {Green}},\
  }\href {\doibase 10.1103/PhysRevB.61.10267} {\bibfield  {journal} {\bibinfo
  {journal} {Phys. Rev. B}\ }\textbf {\bibinfo {volume} {61}},\ \bibinfo
  {pages} {10267} (\bibinfo {year} {2000})} \BibitemShut
  {NoStop}%
\bibitem [{\citenamefont {Fu}\ and\ \citenamefont
  {Kane}(2008)}]{PhysRevLett.100.096407}%
  \BibitemOpen
  \bibfield  {author} {\bibinfo {author} {\bibfnamefont {L.}~\bibnamefont
  {Fu}}\ and\ \bibinfo {author} {\bibfnamefont {C.~L.}\ \bibnamefont {Kane}},\
  }\href {\doibase 10.1103/PhysRevLett.100.096407} {\bibfield  {journal}
  {\bibinfo  {journal} {Phys. Rev. Lett.}\ }\textbf {\bibinfo {volume} {100}},\
  \bibinfo {pages} {096407} (\bibinfo {year} {2008})} \BibitemShut {NoStop}%
\bibitem [{\citenamefont {Qi}\ \emph {et~al.}(2010)\citenamefont {Qi},
  \citenamefont {Hughes},\ and\ \citenamefont {Zhang}}]{PhysRevB.82.184516}%
  \BibitemOpen
  \bibfield  {author} {\bibinfo {author} {\bibfnamefont {X.-L.}\ \bibnamefont
  {Qi}}, \bibinfo {author} {\bibfnamefont {T.~L.}\ \bibnamefont {Hughes}}, \
  and\ \bibinfo {author} {\bibfnamefont {S.-C.}\ \bibnamefont {Zhang}},\ }\href
  {\doibase 10.1103/PhysRevB.82.184516} {\bibfield  {journal} {\bibinfo
  {journal} {Phys. Rev. B}\ }\textbf {\bibinfo {volume} {82}},\ \bibinfo
  {pages} {184516} (\bibinfo {year} {2010})} \BibitemShut {NoStop}%
\bibitem [{\citenamefont {Zhang}\ \emph {et~al.}(2008)\citenamefont {Zhang},
  \citenamefont {Tewari}, \citenamefont {Lutchyn},\ and\ \citenamefont
  {Das~Sarma}}]{PhysRevLett.101.160401}%
  \BibitemOpen
  \bibfield  {author} {\bibinfo {author} {\bibfnamefont {C.}~\bibnamefont
  {Zhang}}, \bibinfo {author} {\bibfnamefont {S.}~\bibnamefont {Tewari}},
  \bibinfo {author} {\bibfnamefont {R.~M.}\ \bibnamefont {Lutchyn}}, \ and\
  \bibinfo {author} {\bibfnamefont {S.}~\bibnamefont {Das~Sarma}},\ }\href
  {\doibase 10.1103/PhysRevLett.101.160401} {\bibfield  {journal} {\bibinfo
  {journal} {Phys. Rev. Lett.}\ }\textbf {\bibinfo {volume} {101}},\ \bibinfo
  {pages} {160401} (\bibinfo {year} {2008})} \BibitemShut {NoStop}%
\bibitem [{\citenamefont {Sato}\ \emph {et~al.}(2009)\citenamefont {Sato},
  \citenamefont {Takahashi},\ and\ \citenamefont
  {Fujimoto}}]{PhysRevLett.103.020401}%
  \BibitemOpen
  \bibfield  {author} {\bibinfo {author} {\bibfnamefont {M.}~\bibnamefont
  {Sato}}, \bibinfo {author} {\bibfnamefont {Y.}~\bibnamefont {Takahashi}}, \
  and\ \bibinfo {author} {\bibfnamefont {S.}~\bibnamefont {Fujimoto}},\ }\href
  {\doibase 10.1103/PhysRevLett.103.020401} {\bibfield  {journal} {\bibinfo
  {journal} {Phys. Rev. Lett.}\ }\textbf {\bibinfo {volume} {103}},\ \bibinfo
  {pages} {020401} (\bibinfo {year} {2009})} \BibitemShut {NoStop}%
\bibitem [{\citenamefont {Sau}\ \emph {et~al.}(2010)\citenamefont {Sau},
  \citenamefont {Lutchyn}, \citenamefont {Tewari},\ and\ \citenamefont
  {Das~Sarma}}]{PhysRevLett.104.040502}%
  \BibitemOpen
  \bibfield  {author} {\bibinfo {author} {\bibfnamefont {J.~D.}\ \bibnamefont
  {Sau}}, \bibinfo {author} {\bibfnamefont {R.~M.}\ \bibnamefont {Lutchyn}},
  \bibinfo {author} {\bibfnamefont {S.}~\bibnamefont {Tewari}}, \ and\ \bibinfo
  {author} {\bibfnamefont {S.}~\bibnamefont {Das~Sarma}},\ }\href {\doibase
  10.1103/PhysRevLett.104.040502} {\bibfield  {journal} {\bibinfo  {journal}
  {Phys. Rev. Lett.}\ }\textbf {\bibinfo {volume} {104}},\ \bibinfo {pages}
  {040502} (\bibinfo {year} {2010})} \BibitemShut {NoStop}%
\bibitem [{\citenamefont {Alicea}(2010)}]{PhysRevB.81.125318}%
  \BibitemOpen
  \bibfield  {author} {\bibinfo {author} {\bibfnamefont {J.}~\bibnamefont
  {Alicea}},\ }\href {\doibase 10.1103/PhysRevB.81.125318} {\bibfield
  {journal} {\bibinfo  {journal} {Phys. Rev. B}\ }\textbf {\bibinfo {volume}
  {81}},\ \bibinfo {pages} {125318} (\bibinfo {year} {2010})} \BibitemShut {NoStop}%
\bibitem [{\citenamefont {Lutchyn}\ \emph {et~al.}(2010)\citenamefont
  {Lutchyn}, \citenamefont {Sau},\ and\ \citenamefont
  {Das~Sarma}}]{PhysRevLett.105.077001}%
  \BibitemOpen
  \bibfield  {author} {\bibinfo {author} {\bibfnamefont {R.~M.}\ \bibnamefont
  {Lutchyn}}, \bibinfo {author} {\bibfnamefont {J.~D.}\ \bibnamefont {Sau}}, \
  and\ \bibinfo {author} {\bibfnamefont {S.}~\bibnamefont {Das~Sarma}},\ }\href
  {\doibase 10.1103/PhysRevLett.105.077001} {\bibfield  {journal} {\bibinfo
  {journal} {Phys. Rev. Lett.}\ }\textbf {\bibinfo {volume} {105}},\ \bibinfo
  {pages} {077001} (\bibinfo {year} {2010})} \BibitemShut {NoStop}%
\bibitem [{\citenamefont {Oreg}\ \emph {et~al.}(2010)\citenamefont {Oreg},
  \citenamefont {Refael},\ and\ \citenamefont {von
  Oppen}}]{PhysRevLett.105.177002}%
  \BibitemOpen
  \bibfield  {author} {\bibinfo {author} {\bibfnamefont {Y.}~\bibnamefont
  {Oreg}}, \bibinfo {author} {\bibfnamefont {G.}~\bibnamefont {Refael}}, \ and\
  \bibinfo {author} {\bibfnamefont {F.}~\bibnamefont {von Oppen}},\ }\href
  {\doibase 10.1103/PhysRevLett.105.177002} {\bibfield  {journal} {\bibinfo
  {journal} {Phys. Rev. Lett.}\ }\textbf {\bibinfo {volume} {105}},\ \bibinfo
  {pages} {177002} (\bibinfo {year} {2010})} \BibitemShut {NoStop}%
\bibitem [{\citenamefont {Mourik}\ \emph {et~al.}(2012)\citenamefont {Mourik},
  \citenamefont {Zuo}, \citenamefont {Frolov}, \citenamefont {Plissard},
  \citenamefont {Bakkers},\ and\ \citenamefont {Kouwenhoven}}]{Mourik2012}%
  \BibitemOpen
  \bibfield  {author} {\bibinfo {author} {\bibfnamefont {V.}~\bibnamefont
  {Mourik}}, \bibinfo {author} {\bibfnamefont {K.}~\bibnamefont {Zuo}},
  \bibinfo {author} {\bibfnamefont {S.~M.}\ \bibnamefont {Frolov}}, \bibinfo
  {author} {\bibfnamefont {S.~R.}\ \bibnamefont {Plissard}}, \bibinfo {author}
  {\bibfnamefont {E.~P. A.~M.}\ \bibnamefont {Bakkers}}, \ and\ \bibinfo
  {author} {\bibfnamefont {L.~P.}\ \bibnamefont {Kouwenhoven}},\ }\href
  {\doibase 10.1126/science.1222360} {\bibfield  {journal} {\bibinfo  {journal}
  {Science}\ }\textbf {\bibinfo {volume} {336}},\ \bibinfo {pages} {1003}
  (\bibinfo {year} {2012})} \BibitemShut {NoStop}%
\bibitem [{\citenamefont {Rokhinson}\ \emph {et~al.}(2012)\citenamefont
  {Rokhinson}, \citenamefont {Liu},\ and\ \citenamefont
  {Furdyna}}]{Rokhinson2012}%
  \BibitemOpen
  \bibfield  {author} {\bibinfo {author} {\bibfnamefont {L.~P.}\ \bibnamefont
  {Rokhinson}}, \bibinfo {author} {\bibfnamefont {X.}~\bibnamefont {Liu}}, \
  and\ \bibinfo {author} {\bibfnamefont {J.~K.}\ \bibnamefont {Furdyna}},\
  }\href {\doibase 10.1038/nphys2429} {\bibfield  {journal} {\bibinfo
  {journal} {Nature Phys.}\ }\textbf {\bibinfo {volume} {8}},\ \bibinfo {pages}
  {795} (\bibinfo {year} {2012})} \BibitemShut {NoStop}%
\bibitem [{\citenamefont {Das}\ \emph {et~al.}(2012)\citenamefont {Das},
  \citenamefont {Ronen}, \citenamefont {Most}, \citenamefont {Oreg},
  \citenamefont {Heiblum},\ and\ \citenamefont {Shtrikman}}]{Das2012}%
  \BibitemOpen
  \bibfield  {author} {\bibinfo {author} {\bibfnamefont {A.}~\bibnamefont
  {Das}}, \bibinfo {author} {\bibfnamefont {Y.}~\bibnamefont {Ronen}}, \bibinfo
  {author} {\bibfnamefont {Y.}~\bibnamefont {Most}}, \bibinfo {author}
  {\bibfnamefont {Y.}~\bibnamefont {Oreg}}, \bibinfo {author} {\bibfnamefont
  {M.}~\bibnamefont {Heiblum}}, \ and\ \bibinfo {author} {\bibfnamefont
  {H.}~\bibnamefont {Shtrikman}},\ }\href {\doibase 10.1038/nphys2479}
  {\bibfield  {journal} {\bibinfo  {journal} {Nature Phys.}\ }\textbf {\bibinfo
  {volume} {8}},\ \bibinfo {pages} {887} (\bibinfo {year} {2012})} \BibitemShut {NoStop}%
\bibitem [{\citenamefont {Deng}\ \emph {et~al.}(2012)\citenamefont {Deng},
  \citenamefont {Yu}, \citenamefont {Huang}, \citenamefont {Larsson},
  \citenamefont {Caroff},\ and\ \citenamefont {Xu}}]{Deng2012}%
  \BibitemOpen
  \bibfield  {author} {\bibinfo {author} {\bibfnamefont {M.~T.}\ \bibnamefont
  {Deng}}, \bibinfo {author} {\bibfnamefont {C.~L.}\ \bibnamefont {Yu}},
  \bibinfo {author} {\bibfnamefont {G.~Y.}\ \bibnamefont {Huang}}, \bibinfo
  {author} {\bibfnamefont {M.}~\bibnamefont {Larsson}}, \bibinfo {author}
  {\bibfnamefont {P.}~\bibnamefont {Caroff}}, \ and\ \bibinfo {author}
  {\bibfnamefont {H.~Q.}\ \bibnamefont {Xu}},\ }\href {\doibase
  10.1021/nl303758w} {\bibfield  {journal} {\bibinfo  {journal} {Nano Lett.}\
  }\textbf {\bibinfo {volume} {12}},\ \bibinfo {pages} {6414} (\bibinfo {year}
  {2012})}\BibitemShut {NoStop}%
\bibitem [{\citenamefont {Ishikawa}(1977)}]{Ishikawa01061977}%
  \BibitemOpen
  \bibfield  {author} {\bibinfo {author} {\bibfnamefont {M.}~\bibnamefont
  {Ishikawa}},\ }\href {\doibase 10.1143/PTP.57.1836} {\bibfield  {journal}
  {\bibinfo  {journal} {Prog. Theor. Phys.}\ }\textbf {\bibinfo {volume}
  {57}},\ \bibinfo {pages} {1836} (\bibinfo {year} {1977})}\BibitemShut
  {NoStop}%
\bibitem [{\citenamefont {Ishikawa}\ \emph {et~al.}(1980)\citenamefont
  {Ishikawa}, \citenamefont {Miyake},\ and\ \citenamefont
  {Usui}}]{Ishikawa01041980}%
  \BibitemOpen
  \bibfield  {author} {\bibinfo {author} {\bibfnamefont {M.}~\bibnamefont
  {Ishikawa}}, \bibinfo {author} {\bibfnamefont {K.}~\bibnamefont {Miyake}}, \
  and\ \bibinfo {author} {\bibfnamefont {T.}~\bibnamefont {Usui}},\ }\href
  {\doibase 10.1143/PTP.63.1083} {\bibfield  {journal} {\bibinfo  {journal}
  {Prog. Theor. Phys.}\ }\textbf {\bibinfo {volume} {63}},\ \bibinfo {pages}
  {1083} (\bibinfo {year} {1980})}\BibitemShut {NoStop}%
\bibitem [{\citenamefont {Mermin}\ and\ \citenamefont
  {Muzikar}(1980)}]{PhysRevB.21.980}%
  \BibitemOpen
  \bibfield  {author} {\bibinfo {author} {\bibfnamefont {N.~D.}\ \bibnamefont
  {Mermin}}\ and\ \bibinfo {author} {\bibfnamefont {P.}~\bibnamefont
  {Muzikar}},\ }\href {\doibase 10.1103/PhysRevB.21.980} {\bibfield  {journal}
  {\bibinfo  {journal} {Phys. Rev. B}\ }\textbf {\bibinfo {volume} {21}},\
  \bibinfo {pages} {980} (\bibinfo {year} {1980})}\BibitemShut {NoStop}%
\bibitem [{\citenamefont {Volovik}(1995)}]{volovik1995}%
  \BibitemOpen
  \bibfield  {author} {\bibinfo {author} {\bibfnamefont {G.~E.}\ \bibnamefont
  {Volovik}},\ }\href@noop {} {\bibfield  {journal} {\bibinfo  {journal} {JETP
  Lett.}\ }\textbf {\bibinfo {volume} {61}},\ \bibinfo {pages} {958} (\bibinfo
  {year} {1995})}\BibitemShut {NoStop}%
\bibitem [{\citenamefont {Kita}(1998)}]{JPSJ.67.216}%
  \BibitemOpen
  \bibfield  {author} {\bibinfo {author} {\bibfnamefont {T.}~\bibnamefont
  {Kita}},\ }\href {\doibase 10.1143/JPSJ.67.216} {\bibfield  {journal}
  {\bibinfo  {journal} {J. Phys. Soc. Jpn.}\ }\textbf {\bibinfo {volume}
  {67}},\ \bibinfo {pages} {216} (\bibinfo {year} {1998})}\BibitemShut
  {NoStop}%
\bibitem [{\citenamefont {Goryo}\ and\ \citenamefont
  {Ishikawa}(1998)}]{Goryo1998549}%
  \BibitemOpen
  \bibfield  {author} {\bibinfo {author} {\bibfnamefont {J.}~\bibnamefont
  {Goryo}}\ and\ \bibinfo {author} {\bibfnamefont {K.}~\bibnamefont
  {Ishikawa}},\ }\href {\doibase 10.1016/S0375-9601(98)00438-1} {\bibfield
  {journal} {\bibinfo  {journal} {Phys. Lett. A}\ }\textbf {\bibinfo {volume}
  {246}},\ \bibinfo {pages} {549 } (\bibinfo {year} {1998})} \BibitemShut
  {NoStop}%
\bibitem [{\citenamefont {Stone}\ and\ \citenamefont
  {Roy}(2004)}]{PhysRevB.69.184511}%
  \BibitemOpen
  \bibfield  {author} {\bibinfo {author} {\bibfnamefont {M.}~\bibnamefont
  {Stone}}\ and\ \bibinfo {author} {\bibfnamefont {R.}~\bibnamefont {Roy}},\
  }\href {\doibase 10.1103/PhysRevB.69.184511} {\bibfield  {journal} {\bibinfo
  {journal} {Phys. Rev. B}\ }\textbf {\bibinfo {volume} {69}},\ \bibinfo
  {pages} {184511} (\bibinfo {year} {2004})} \BibitemShut
  {NoStop}%
\bibitem [{\citenamefont {Sauls}(2011)}]{PhysRevB.84.214509}%
  \BibitemOpen
  \bibfield  {author} {\bibinfo {author} {\bibfnamefont {J.~A.}\ \bibnamefont
  {Sauls}},\ }\href {\doibase 10.1103/PhysRevB.84.214509} {\bibfield  {journal}
  {\bibinfo  {journal} {Phys. Rev. B}\ }\textbf {\bibinfo {volume} {84}},\
  \bibinfo {pages} {214509} (\bibinfo {year} {2011})} \BibitemShut {NoStop}%
\bibitem [{\citenamefont {Tsutsumi}\ and\ \citenamefont
  {Machida}(2012)}]{PhysRevB.85.100506}%
  \BibitemOpen
  \bibfield  {author} {\bibinfo {author} {\bibfnamefont {Y.}~\bibnamefont
  {Tsutsumi}}\ and\ \bibinfo {author} {\bibfnamefont {K.}~\bibnamefont
  {Machida}},\ }\href {\doibase 10.1103/PhysRevB.85.100506} {\bibfield
  {journal} {\bibinfo  {journal} {Phys. Rev. B}\ }\textbf {\bibinfo {volume}
  {85}},\ \bibinfo {pages} {100506} (\bibinfo {year} {2012})} \BibitemShut {NoStop}%
\bibitem [{\citenamefont {Shitade}\ and\ \citenamefont
  {Kimura}(2014)}]{PhysRevB.90.134510}%
  \BibitemOpen
  \bibfield  {author} {\bibinfo {author} {\bibfnamefont {A.}~\bibnamefont
  {Shitade}}\ and\ \bibinfo {author} {\bibfnamefont {T.}~\bibnamefont
  {Kimura}},\ }\href {\doibase 10.1103/PhysRevB.90.134510} {\bibfield
  {journal} {\bibinfo  {journal} {Phys. Rev. B}\ }\textbf {\bibinfo {volume}
  {90}},\ \bibinfo {pages} {134510} (\bibinfo {year} {2014})} \BibitemShut {NoStop}%
\bibitem [{\citenamefont {Mizushima}\ \emph {et~al.}(2008)\citenamefont
  {Mizushima}, \citenamefont {Ichioka},\ and\ \citenamefont
  {Machida}}]{PhysRevLett.101.150409}%
  \BibitemOpen
  \bibfield  {author} {\bibinfo {author} {\bibfnamefont {T.}~\bibnamefont
  {Mizushima}}, \bibinfo {author} {\bibfnamefont {M.}~\bibnamefont {Ichioka}},
  \ and\ \bibinfo {author} {\bibfnamefont {K.}~\bibnamefont {Machida}},\ }\href
  {\doibase 10.1103/PhysRevLett.101.150409} {\bibfield  {journal} {\bibinfo
  {journal} {Phys. Rev. Lett.}\ }\textbf {\bibinfo {volume} {101}},\ \bibinfo
  {pages} {150409} (\bibinfo {year} {2008})} \BibitemShut {NoStop}%
\bibitem [{\citenamefont {Mizushima}\ and\ \citenamefont
  {Machida}(2010)}]{PhysRevA.81.053605}%
  \BibitemOpen
  \bibfield  {author} {\bibinfo {author} {\bibfnamefont {T.}~\bibnamefont
  {Mizushima}}\ and\ \bibinfo {author} {\bibfnamefont {K.}~\bibnamefont
  {Machida}},\ }\href {\doibase 10.1103/PhysRevA.81.053605} {\bibfield
  {journal} {\bibinfo  {journal} {Phys. Rev. A}\ }\textbf {\bibinfo {volume}
  {81}},\ \bibinfo {pages} {053605} (\bibinfo {year} {2010})} \BibitemShut {NoStop}%
\bibitem [{\citenamefont {Tada}\ \emph {et~al.}(2015)\citenamefont {Tada},
  \citenamefont {Nie},\ and\ \citenamefont
  {Oshikawa}}]{PhysRevLett.114.195301}%
  \BibitemOpen
  \bibfield  {author} {\bibinfo {author} {\bibfnamefont {Y.}~\bibnamefont
  {Tada}}, \bibinfo {author} {\bibfnamefont {W.}~\bibnamefont {Nie}}, \ and\
  \bibinfo {author} {\bibfnamefont {M.}~\bibnamefont {Oshikawa}},\ }\href
  {\doibase 10.1103/PhysRevLett.114.195301} {\bibfield  {journal} {\bibinfo
  {journal} {Phys. Rev. Lett.}\ }\textbf {\bibinfo {volume} {114}},\ \bibinfo
  {pages} {195301} (\bibinfo {year} {2015})} \BibitemShut {NoStop}%
\bibitem [{\citenamefont {Volovik}(2014)}]{Volovik2014}%
  \BibitemOpen
  \bibfield  {author} {\bibinfo {author} {\bibfnamefont {G.~E.}\ \bibnamefont
  {Volovik}},\ }\href@noop {} {\bibfield  {journal} {\bibinfo  {journal} {JETP
  Lett.}\ }\textbf {\bibinfo {volume} {100}},\ \bibinfo {pages} {742} (\bibinfo
  {year} {2014})} \BibitemShut {NoStop}%
\bibitem [{\citenamefont {Huang}\ \emph {et~al.}(2014)\citenamefont {Huang},
  \citenamefont {Taylor},\ and\ \citenamefont {Kallin}}]{PhysRevB.90.224519}%
  \BibitemOpen
  \bibfield  {author} {\bibinfo {author} {\bibfnamefont {W.}~\bibnamefont
  {Huang}}, \bibinfo {author} {\bibfnamefont {E.}~\bibnamefont {Taylor}}, \
  and\ \bibinfo {author} {\bibfnamefont {C.}~\bibnamefont {Kallin}},\ }\href
  {\doibase 10.1103/PhysRevB.90.224519} {\bibfield  {journal} {\bibinfo
  {journal} {Phys. Rev. B}\ }\textbf {\bibinfo {volume} {90}},\ \bibinfo
  {pages} {224519} (\bibinfo {year} {2014})} \BibitemShut {NoStop}%
\bibitem [{\citenamefont {Hu}\ \emph {et~al.}(2013)\citenamefont {Hu},
  \citenamefont {Jiang}, \citenamefont {Pu}, \citenamefont {Chen},\ and\
  \citenamefont {Liu}}]{PhysRevLett.110.020401}%
  \BibitemOpen
  \bibfield  {author} {\bibinfo {author} {\bibfnamefont {H.}~\bibnamefont
  {Hu}}, \bibinfo {author} {\bibfnamefont {L.}~\bibnamefont {Jiang}}, \bibinfo
  {author} {\bibfnamefont {H.}~\bibnamefont {Pu}}, \bibinfo {author}
  {\bibfnamefont {Y.}~\bibnamefont {Chen}}, \ and\ \bibinfo {author}
  {\bibfnamefont {X.-J.}\ \bibnamefont {Liu}},\ }\href {\doibase
  10.1103/PhysRevLett.110.020401} {\bibfield  {journal} {\bibinfo  {journal}
  {Phys. Rev. Lett.}\ }\textbf {\bibinfo {volume} {110}},\ \bibinfo {pages}
  {020401} (\bibinfo {year} {2013})} \BibitemShut {NoStop}%
\bibitem [{\citenamefont {Nagai}\ \emph
  {et~al.}(2014{\natexlab{a}})\citenamefont {Nagai}, \citenamefont {Ota},\ and\
  \citenamefont {Machida}}]{JPSJ.83.094722}%
  \BibitemOpen
  \bibfield  {author} {\bibinfo {author} {\bibfnamefont {Y.}~\bibnamefont
  {Nagai}}, \bibinfo {author} {\bibfnamefont {Y.}~\bibnamefont {Ota}}, \ and\
  \bibinfo {author} {\bibfnamefont {M.}~\bibnamefont {Machida}},\ }\href
  {\doibase 10.7566/JPSJ.83.094722} {\bibfield  {journal} {\bibinfo  {journal}
  {J.Phys. Soc. Jpn.}\ }\textbf {\bibinfo {volume} {83}},\ \bibinfo {pages}
  {094722} (\bibinfo {year} {2014}{\natexlab{a}})} \BibitemShut {NoStop}%
\bibitem [{\citenamefont {Nagai}\ \emph {et~al.}(2015)\citenamefont {Nagai},
  \citenamefont {Ota},\ and\ \citenamefont {Machida}}]{JPSJ.84.034711}%
  \BibitemOpen
  \bibfield  {author} {\bibinfo {author} {\bibfnamefont {Y.}~\bibnamefont
  {Nagai}}, \bibinfo {author} {\bibfnamefont {Y.}~\bibnamefont {Ota}}, \ and\
  \bibinfo {author} {\bibfnamefont {M.}~\bibnamefont {Machida}},\ }\href
  {\doibase 10.7566/JPSJ.84.034711} {\bibfield  {journal} {\bibinfo  {journal}
  {J. Phys. Soc. Jpn.}\ }\textbf {\bibinfo {volume} {84}},\ \bibinfo {pages}
  {034711} (\bibinfo {year} {2015})} \BibitemShut {NoStop}%
\bibitem [{\citenamefont {Masaki}\ and\ \citenamefont
  {Kato}(2014)}]{1742-6596-568-2-022028}%
  \BibitemOpen
  \bibfield  {author} {\bibinfo {author} {\bibfnamefont {Y.}~\bibnamefont
  {Masaki}}\ and\ \bibinfo {author} {\bibfnamefont {Y.}~\bibnamefont {Kato}},\
  }\href {\doibase 10.1088/1742-6596/568/2/022028} {\bibfield  {journal}
  {\bibinfo  {journal} {J. Phys.: Conf. Ser.}\ }\textbf {\bibinfo {volume}
  {568}},\ \bibinfo {pages} {022028} (\bibinfo {year} {2014})}\BibitemShut
  {NoStop}%
\bibitem [{\citenamefont {Read}(2009)}]{PhysRevB.79.045308}%
  \BibitemOpen
  \bibfield  {author} {\bibinfo {author} {\bibfnamefont {N.}~\bibnamefont
  {Read}},\ }\href {\doibase 10.1103/PhysRevB.79.045308} {\bibfield  {journal}
  {\bibinfo  {journal} {Phys. Rev. B}\ }\textbf {\bibinfo {volume} {79}},\
  \bibinfo {pages} {045308} (\bibinfo {year} {2009})} \BibitemShut {NoStop}%
\bibitem [{\citenamefont {Read}\ and\ \citenamefont
  {Rezayi}(2011)}]{PhysRevB.84.085316}%
  \BibitemOpen
  \bibfield  {author} {\bibinfo {author} {\bibfnamefont {N.}~\bibnamefont
  {Read}}\ and\ \bibinfo {author} {\bibfnamefont {E.~H.}\ \bibnamefont
  {Rezayi}},\ }\href {\doibase 10.1103/PhysRevB.84.085316} {\bibfield
  {journal} {\bibinfo  {journal} {Phys. Rev. B}\ }\textbf {\bibinfo {volume}
  {84}},\ \bibinfo {pages} {085316} (\bibinfo {year} {2011})} \BibitemShut {NoStop}%
\end{thebibliography}
\end{document}